\documentclass[aps, showpacs, pra, superscriptadaddress, 
twocolumn] {revtex4-1}
\usepackage{hyperref}
 \usepackage{amsmath}
\usepackage{amssymb}
\usepackage{epsfig}
\usepackage{natbib}
\usepackage{epstopdf}
\usepackage{graphicx}

\usepackage{floatrow}
\usepackage[caption=false,font=Large]{subfig}

\newcommand*{\be}{\begin{equation}}
\newcommand*{\ee}{\end{equation}}
\newcommand*{\bea}{\begin{eqnarray}}
\newcommand*{\eea}{\end{eqnarray}}

\newcommand{\PT}{$\mathcal{PT}$}

 \DeclareFontFamily{OT1}{pzc}{}
 \DeclareFontShape{OT1}{pzc}{m}{it}%
 {<->  s  *  [1.400]  pzcmi7t}{}
\DeclareMathAlphabet{\mathscr}{OT1}{pzc}%
{m}{it}

\begin{document}

\title{Integrability and trajectory confinement in \PT-symmetric waveguide arrays}

\author{I V Barashenkov and Frank Smuts}    
\affiliation{
Centre for Theoretical and Mathematical Physics,  University of Cape Town,  South Africa}

 
 \author{Alexander Chernyavsky} 
 \affiliation{
 Department of Mathematics, State University of New York at Buffalo,  Buffalo, NY 14260 USA
} 
  
\begin{abstract} 
We consider \PT-symmetric ring-like arrays of optical waveguides  with purely nonlinear 
 gain and loss. Regardless of the value of the gain-loss coefficient,
 these systems are protected
 from spontaneous \PT-symmetry breaking. 
  If the nonhermitian part of the array matrix   has cross-compensating  structure, 
  the total power in such a system remains bounded --- or even constant --- at all times. 
  We identify two-, three-, and four-waveguide arrays with cross-compensatory nonlinear gain and loss
that constitute completely integrable Hamiltonian systems. 
\end{abstract}

\maketitle

\section{Introduction}

The concept of $\mathcal{PT}$ symmetry,
originally introduced in nonhermitian quantum mechanics
\cite{Bender,Bender_book},  has led to significant developments in   photonics,  plasmonics,  quantum optics of atomic gases, 
metamaterials, Bose-Einstein condensates, electronic circuitry, and acoustics 
\cite{Bender_book,reviews}.
The $\mathcal{PT}$-symmetric equations 
model physical structures with a built-in  balance between gain and loss.

In a linear nonhermitian system, raising the gain-loss coefficient above a critical level causes  trajectories  to escape to infinity. 
The setting in of this blow-up  instability can have a detrimental effect on the  structure ---  in optics, for example, an escaping trajectory implies an uncontrollable power hike.

Taking into account nonlinear effects and adding nonlinear corrections to the equations may arrest the blow-up  through  the emergence of conserved quantities confining trajectories 
to a finite part of the phase space \cite{Dubard}.
In this paper, we study a class of \PT-symmetric systems with the nonhermiticity induced {\it entirely\/}  by the nonlinear terms. 
On one hand, these systems provide access to the full set of behaviours afforded by the presence of gain and loss. On the other hand, they exhibit remarkable regularity properties such as 
the existence of the Hamiltonian structure and  trajectory confinement.

The Hamiltonian structure in a dynamical system imposes a deep symmetry between two sets of coordinates parameterising its phase space. 
In the presence of additional first integrals, the Hamiltonian structure establishes an even higher degree of regularity:  the Liouville integrability.
We identify  two-, three- and four-component 
 integrable systems with nonhermitian nonlinearities that are free from the blow-up behaviour.

The two-component complex systems that  we study have the form of the nonlinear Schr\"odinger dimer: a  discrete Schr\"odinger equation, defined on only two sites  \cite{Hamilt_dimer,I1,dimer,Tyug_Susanto,Jackson,standard}.
The nonhermitian dimer serves  as an archetypal model for a pair of  optical waveguides or a pair of micro-ring resonators with gain and loss, 
coupled by their evanescent fields    \cite{waveguides,Miri}.
It also arises in the study of  Bose-Einstein condensates \cite{Graefe,BEC}, plasmonics \cite{plasmonics}, spintronics \cite{BC}, 
electronic circuitry \cite{electronics} 
and several other contexts.

The literature suggests several recipes for  preventing 
the blow-up in dimers, including linear vs nonlinear gain-loss competition \cite{lin_nonlin,Karthiga} and nonlinear gain-loss saturation \cite{Miri,Huerta}.
This paper explores cross-stimulation --- an alternative mechanism that, in addition to ensuring nonsingular evolution, conserves the norm $|u|^2+|v|^2$
(interpreted as the 
total power of light in the optical context).
 We present two cross-stimulated \PT-symmetric dimers that describe completely integrable Hamiltonian systems.

The  cross-stimulation is a special type of a more general notion of cross-compensation of gain and loss
in a multichannel structure.   To illustrate  this concept,   we invoke a
 three- and four-site discrete Schr\"odinger equation --- the \PT-symmetric trimer and quadrimer, respectively
  \cite{trimer}. 
      In the optical domain, the trimer and quadrimer 
model  an optical  necklace --- an 
array of three or four coupled waveguides or resonators.
We produce examples of a completely integrable   \PT-symmetric trimer and quadrimer, with  solutions free from the blow-up behaviour.
Similar to the dimers in the first part of this study, the nonherimiticity of these necklaces is entirely due to the nonlinear terms
and has a cross-compensatory character.

The paper is organised into five sections. We start with a linear hermitian dimer with the cross-stimulating cubic gain and loss
(section \ref{lindim}). The subsequent section 
(section \ref{nlin_cross})  deals with a slightly more complex cross-stimulating system whose hermitian part  is cubic itself.
We uncover the hidden Hamiltonian structure of these systems, determine their integrals of motion and construct analytic solutions. 
 An integrable \PT-symmetric trimer and quadrimer  are identified in section \ref{necks}.
Section \ref{Conclusions} summarises results of this study.

\section{Linear dimer with nonlinear gain and loss}
\label{lindim}
\subsection{Gain and loss cross-stimulation} 

As the gain-loss coefficient is varied,  the topological structure of the phase portrait 
 of a  dimer
with linear gain and loss undergoes a spontaneous change. 
 Consider, for example, 
the so-called {\it standard\/} \PT-symmetric dimer,  a model that arises in a 
wide range of physical contexts  \cite{standard,Tyug_Susanto,Jackson,I1}:
\begin{subequations} \label{sta}
\begin{align}
    iu_t + v  +|u|^2u &= i \gamma  u,   \\
    iv_t +u +|v|^2v   &= - i \gamma  v.   
\end{align}   \end{subequations} 
As    $\gamma$ is raised above the critical value $\gamma_c =1$, 
the fixed point at $u=v=0$ loses its stability and the \PT-symmetry is said to become spontaneously broken. 
In the symmetry-broken phase ($\gamma>1$) small initial conditions give rise to exponentially
growing solutions. Since the change of behaviour concerns solutions of the linearised equations, we 
refer to this bifurcation  as the   {\it linearised\/}  symmetry breaking.

In a generic  Schr\"odinger dimer,
 trajectories resulting from the exponentially growing  solutions of the linearised equations 
  may escape to infinity. 
 For example, in the standard dimer 
\eqref{sta} {\it all\/} growing linearised solutions give rise to escaping trajectories \cite{Jackson}. 
(Note that unbounded solutions may occur in the symmetric phase too --- 
they just need to evolve out of large initial data \cite{Tyug_Susanto,Jackson}.)

To  preclude the  blow-up of small initial data, we consider  a system with
a hermitian linearised matrix. 
Furthermore,
our dimer is assumed to be cross-stimulated. This means that 
 the $u$-channel gains energy at the rate proportional 
to the power carried by its $v$-neighbour while 
the $v$ amplitude loses energy at a rate proportional to the power carried by
 $u$:
\begin{subequations} \label{A100}
\begin{align}
    iu_t + v  &= i \gamma |v|^2 u, \label{A5}  \\
    iv_t +u   &= - i \gamma |u|^2 v.   \label{A6}
\end{align}   \end{subequations} 
As a result of the cross-stimulation, 
the total power
$P=|u|^2+|v|^2$ is conserved.  This  keeps {\it all\/} trajectories --- both with small and large initial data ---
in a finite part of the phase space.

Although the cross-stimulation  may come across as a purely mathematical  construct,
 the system \eqref{A100}   originates in a well-established physical context. 
  It describes
 the spin-torque oscillator — an isotropic ferromagnet in an external magnetic field with polarised spin current driven through it \cite{BC}.
The components of the total spin vector $ {\bf S} = \{ X,Y,Z\}$ in the free layer of the oscillator 
are expressible through the complex amplitudes of the dimer's channels:
\begin{equation}
    X = \frac{\bar{u} v + u\bar{v}}{2}, \quad 
    Y = i  \frac{\bar{u} v - u\bar{v}}{2}, \quad 
    Z = \frac{|u|^2 - |v|^2}{2}.
    \label{Stokes-variables}
\end{equation}
The length of the vector, $\mathcal{R} = \sqrt{X^2 + Y^2 + Z^2}$, is proportional to the total power carried by the dimer,
\[
    \mathcal{R} = \frac{ |u|^2 + |v|^2}{2}, 
\]
 which is fixed 
by the initial condition.
The direction of the vector is determined by a system of three equations \cite{BC}:
\begin{subequations}
 \label{1st3}
\begin{align}
    \dot{X} &= -\gamma X Z,    \label{A7} \\
    \dot{Y} &= -(1+\gamma Y) Z,       \label{A8}   \\
    \dot{Z} &= Y + \gamma (R^2 - Z^2),   \label{A9}
\end{align}  \end{subequations} 
where $X,Y$ and $Z$ are considered to be functions of $\tau=2t$, and the overdot stands for the derivative w.r.t. $\tau$.

Another area of applications of  the cross-stimulated dimers is 
 the classical limit of quantum theory.  A
system equivalent to \eqref{A100}
  governs the normalised state vector of the  nonhermitian two-level atom \cite{Graefe}.
  The corresponding Bloch-sphere dynamics  obey equations \eqref{1st3}. 

\subsection{Hamiltonian structure and integrability}

          \begin{figure}[t]
 \begin{center} 
       \includegraphics*[width=0.65\linewidth] {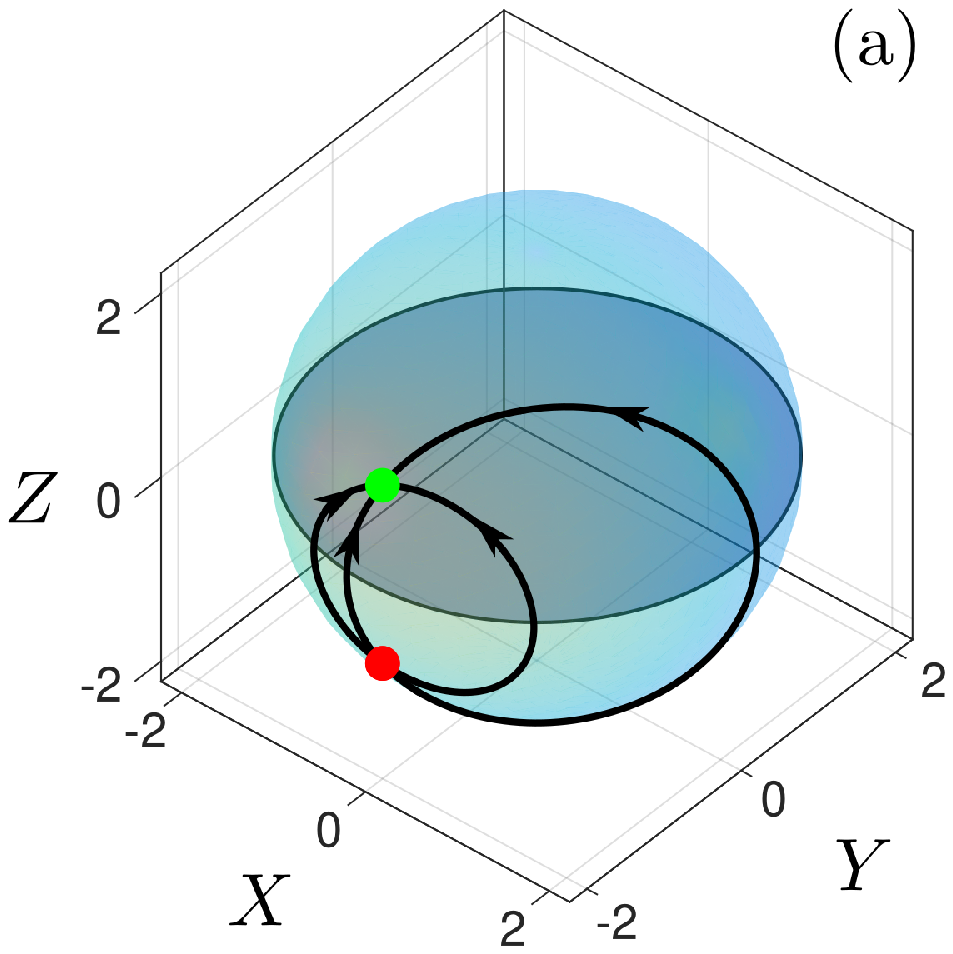}  
   \includegraphics*[width=0.65\linewidth] {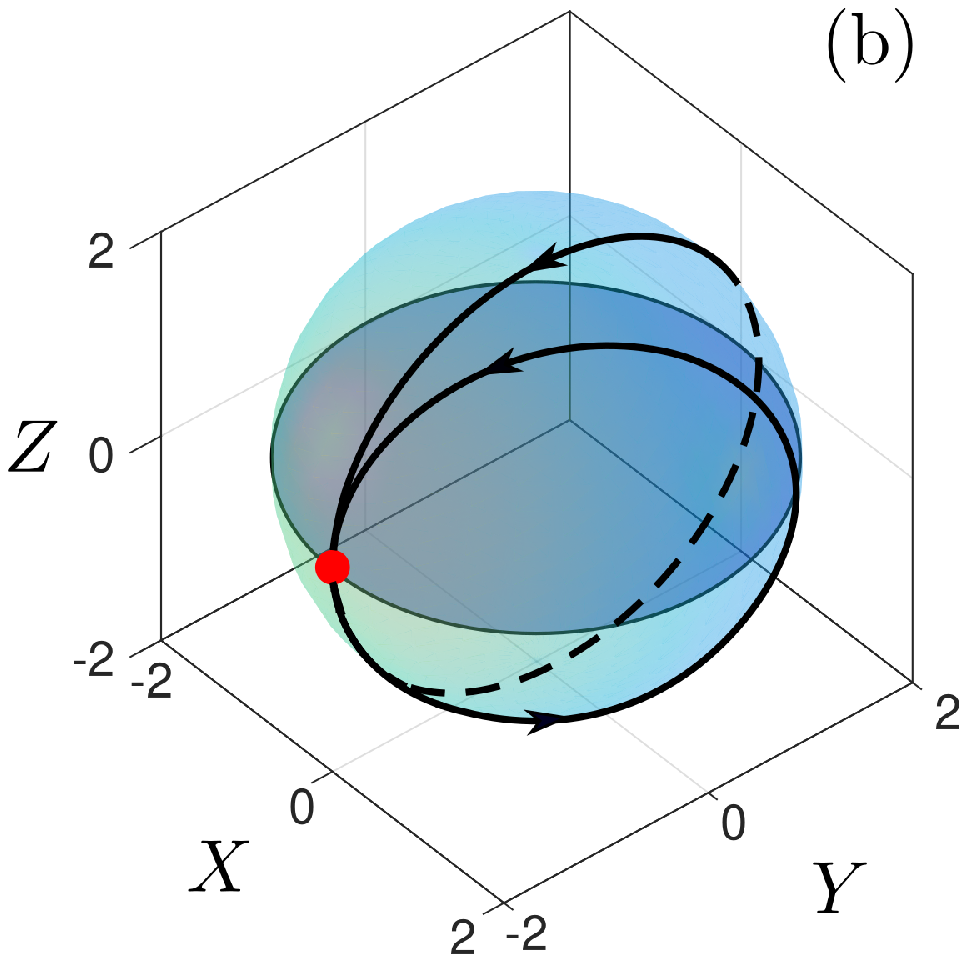}  
    \includegraphics*[width=0.65\linewidth] {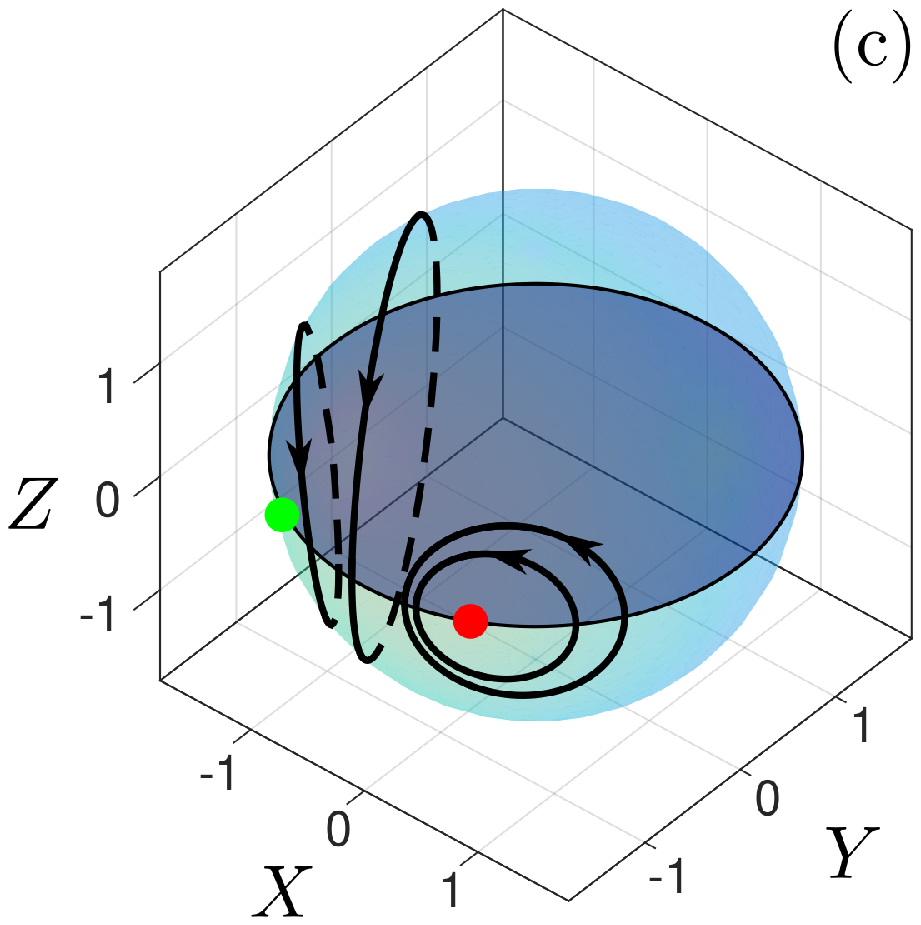}  
              \end{center}
  \caption{Trajectories of the system \eqref{1st3} on the surface of the sphere of the 
  radius  $\mathcal R> \gamma^{-1}$   (a);    $\mathcal R=  \gamma^{-1}$ (b) and
  $\mathcal R<  \gamma^{-1}$ (c). (In this plot, $\gamma=\frac12$.)
  Note that the largest sphere (panel a)   has been scaled down while the smallest one 
  (panel c) has been scaled up.
  In each ball, tinted is the equatorial plane. \label{3figs}
  }
 \end{figure}

Upon   defining a set of polar 
coordinates by
\be
X=e^{-\eta} \cos \theta, \quad
Y=  -\gamma^{-1}+
e^{-\eta} \sin \theta,   \label{polars}
\ee
 equations \eqref{1st3} simplify  to
\begin{align}
{\dot \eta}= \gamma Z,       \quad
{\dot \theta} =0,  \label{A41}  \\
{\dot Z}= \gamma e^{-2\eta} - e^{-\eta} \sin \theta.  \label{A42}
\end{align}   
     It is not difficult to see that the system \eqref{A41}-\eqref{A42}  is Hamiltonian.
     We choose the integral ${\mathcal R}^2-\gamma^{-2}$ as 
   the Hamilton function:
   \be
   H(\eta, \theta, p_\eta)= e^{-2\eta} - 2\gamma^{-1} e^{-\eta} \sin \theta +\frac{\gamma^2}{4} p^2_\eta.
   \label{A25}
   \ee
Here $\eta$ and $\theta$ are the canonical coordinates and $p_\eta= 2\gamma^{-1}Z$ is the canonical momentum
conjugate to $\eta$. The equations 
\[
\dot \eta= \frac{\partial H}{\partial p_\eta},
\quad
\dot \theta= \frac{\partial H}{\partial p_\theta}
\]
reproduce equations \eqref{A41}, and 
\[
\dot p_\eta= - \frac{\partial H}{\partial \eta}
\]
amounts to \eqref{A42}. The last Hamilton equation,
\[
\dot p_\theta= - \frac{\partial H}{\partial \theta},
\]
serves as a definition of the momentum $p_\theta$:
\[
p_\theta= \frac{2}{\gamma} \int_0^\tau X(\tau') d \tau'.
\]

The existence of the Hamiltonian structure and an additional first integral 
 --- the azimuthal angle $\theta$ --- establishes the complete integrability of the 
dimer \eqref{A100}. 

 \subsection{Solutions}

The trajectories of the system  \eqref{1st3}   
are circular arcs  resulting from  the 
section of the sphere $X^2+Y^2+Z^2= \mathcal R^2$ by vertical 
half-planes $\theta=const$. 
These were classified in an earlier study \cite{Graefe} in terms of the fixed points of the system. 
Changing to the canonical  variables \eqref{polars} allows us to express  solutions of the dimer \eqref{A100}  explicitly.
 
 Using \eqref{A25} and the first equation in \eqref{A41}, we obtain
 \be
\dot K^2 - \nu^2 K^2-  2 \gamma \sin (\theta) K + \gamma^2=0,     \label{qua}
\ee 
where $K= e^\eta$ and $\nu^2= \gamma^2 {\mathcal R}^2  -1$.
This is a   conserved quantity of the linear equation 
 \be
\ddot K -  \nu^2 K = \gamma \sin \theta.    \label{2order}
\ee

The sphere with radius $\mathcal R> \gamma^{-1}$ encloses a section of the vertical axis $X=0$, $Y=-\gamma^{-1}$.
Accordingly, the sphere 
 is crossed by vertical half-planes with all possible azimuthal angles,
 from  $\theta=0$   to  $\theta=2 \pi$. The resulting trajectories are circular arcs that 
connect an attracting fixed point (a stable node) at
\[
X=0, \quad Y=-\gamma^{-1}, \quad  Z= - \sqrt{\mathcal R^2-\gamma^{-2}}
\]
 to a repelling  point (unstable node) that has the same $X,Y$ and opposite $Z$  (Fig \ref{3figs}(a)).
The corresponding   solutions of \eqref{2order} are given by
\be
K= Ae^{\nu \tau} +Be^{-\nu \tau} -   \gamma \nu^{-2} \sin \theta, \label{A60} \\
\ee
where the relation between the  constants of integration is found by 
substituting \eqref{A60} into \eqref{qua}:
\[
  AB = \frac{\gamma^2}{4\nu^2} \left( \frac{\sin^2\theta}{\nu^2} + 1 \right).
\]

For each value of  spherical  radius $\mathcal R$
 the  equations  \eqref{1st3} represent  a dynamical system on the $(X,Y)$-plane,
 with a conserved 
quantity $\theta(X,Y)$. 
We note an ostensible paradox, where the existence of the conservation law should be precluded by the presence of the attractor and repeller in the phase space.
The paradox is resolved, however,  upon observing that $\theta$ is 
undefined at either  fixed point.

As $\mathcal R$ is decreased through $\gamma^{-1}$, the attracting and repelling points approach each other, collide  and then diverge along the equator of the sphere.
The sphere with  $\mathcal R< \gamma^{-1}$ has two elliptic fixed points, at 
\[
X=\pm \mathcal R\sqrt{1-\gamma^2 \mathcal R^2}, \quad
Y= -\gamma \mathcal R^2,
\quad 
 Z=0.
 \]
  Each point is surrounded by a family of  circular orbits  (Fig \ref{3figs}(c)).
The corresponding
 solutions of \eqref{2order} are
  periodic:
 \[
K= A \cos  (\omega \tau)+ \gamma \omega^{-2} \sin \theta.
\]
  Here, 
  \be
  \omega= \sqrt{1-\gamma^2 \mathcal{R}^2}, \quad    A = \frac{\gamma}{\omega} \sqrt{ \gamma^2 \mathcal R^2- \cos^2 \theta}.
     \label{AqA}
\ee

The vertical axis $X=0, Y=-\gamma^{-1}$ lies outside the  sphere with $\mathcal R < \gamma^{-1}$. 
As a result, the sphere is only crossed by the half-planes with $\theta$ in the interval defined by 
equation \eqref{AqA}:
  \[
  \arccos (\gamma \mathcal R) < \theta< \pi- \arccos (\gamma \mathcal R).
  \]

In the borderline case $\mathcal R= \gamma^{-1}$, the system    \eqref{1st3}    has a single fixed point, at $X=Z=0$ and $Y=-\gamma^{-1}$. 
Trajectories are homoclinic: they emerge out of the fixed point, wrap around the sphere and flow into the same semistable point (Fig \ref{3figs}(b)).
The corresponding solution of equation \eqref{qua} is 
\[
K= \gamma \sin \theta  \, \frac{\tau^2}{2} +F \tau+G,   
\]
where
 \begin{equation*}
 G= \frac{F^2+\gamma^2}{2 \gamma \sin \theta}.
 \end{equation*}

 Once we have an explicit expression for $X(\tau)$, $Y(\tau)$ and $Z(\tau)$,
  the corresponding dimer
 components  can be easily reconstructed: 
 \begin{align*}
 u= \sqrt{ \mathcal R+Z}  \exp \left\{
  \frac{i}{2}  \int_0^\tau  \frac{X}{\mathcal R+ Z} d \tau' + i  \, \mathrm{Arg} \,  u(0) \right\}, \\
  v= \sqrt{ \mathcal R-Z} \exp \left\{
  \frac{i}{2}  \int_0^\tau  \frac{X}{\mathcal R- Z} d \tau'  + i  \, \mathrm{Arg}  \, v(0)  \right\}.
 \end{align*}

\subsection{Related systems}

We close this section with three remarks. Firstly, the absence of any
conservative nonlinearity in the cross-stimulated dimer 
is not a prerequisite for its integrability.
A simple example of  the model with a  nonlinear hermitian part and cross-stimulated channels is
 \begin{subequations}  \label{A200} 
\begin{align}
    i U_t + V + (|U|^2+|V|^2)   U &= i \gamma |V|^2  U,   
      \\
    i V_t + U + (|U|^2+|V|^2)     V &= - i \gamma |U|^2 V.  
\end{align}   \end{subequations} 
The dimer \eqref{A200} maps onto 
 our system  \eqref{A100} and inherits its integrability and trajectory-confinement property. 
 The
gauge transformation relating the two systems is simply
\[ 
U(t)=e^{-i(|u|^2+|v|^2)t}
u(t),
\quad
 V(t)= 
 e^{-i(|u|^2+|v|^2) t}
v(t).
\]

Secondly,
  the  transformation 
  \begin{subequations}   \label{E1E2} 
\begin{align}
E_1(t)= \exp \left\{  \frac{\gamma}{2} \int_0^t (|u|^2-|v|^2) dt' \right\}  u(t), \\
E_2(t)= \exp  \left\{   \frac{\gamma}{2} \int_0^t (|u|^2-|v|^2) dt'  \right\} v(t)
\end{align}   \end{subequations} 
 takes  a solution of
 the system \eqref{A100} with $|u|^2+|v|^2= 2 \mathcal R$ to a 
 solution of the  linear dimer
 \[
    i \partial_t E_1 + E_2  = i \tilde  \gamma  E_1,         \quad
    i\partial_t E_2 + E_1   = -i  \tilde \gamma  E_2, 
    \]
    with $\tilde \gamma = \gamma \mathcal R$. 
This correspondence accounts for the reducibility of  the cross-stimulated dimer  to 
 a linear equation (equation \eqref{2order}).

Finally, it is appropriate to
 mention an earlier study  \cite{Karthiga}
in which the authors constructed integrals of motion of the  system
\begin{subequations} \label{A500}
\begin{align}
    iu_t + v  &= i (\alpha + \beta |u|^2 +\gamma |v|^2) u,   \\
    iv_t +u   &= - i (\alpha + \beta |v|^2 + \gamma |u|^2) v.
\end{align}   \end{subequations} 
The model \eqref{A500} includes our dimer  \eqref{A100}  as a particular case  with $\alpha=\beta=0$.
However, the  integrals of motion  of  \eqref{A500} do not persist as $\beta \to 0$;
see Ref \cite{Karthiga}.

\section{Kerr dimer with  cross-stimulated gain and loss}
\label{nlin_cross} 

\subsection{The system} 
 
Having outlined the effect  of cross-stimulation on a model dimer \eqref{A100},
we observe that this mechanism remains available to a broad 
class of systems in nonlinear optics. 
If we   retain the generic Kerr nonlinearity
in   equations \eqref{A100}, 
we obtain  another cross-stimulated dimer conserving the total power $|u|^2+|v|^2$:
 \begin{subequations}   \label{A3A4}
\begin{align}
    iu_t + v +  |u|^2u &= i \gamma |v|^2 u, \label{A3}  \\
    iv_t +u  +  |v|^2v &= - i \gamma |u|^2 v.   \label{A4}
\end{align}
\end{subequations}

Equations   \eqref{A3A4}  result from the 
model of a birefringent single-mode fibre amplifier with a saturable nonlinearity \cite{Huerta}:
\begin{align*}
    i \partial_t E_1 + E_2 +  \frac{ 2 \mathcal R  |E_1|^2}{|E_1|^2+|E_2|^2} E_1 &= i \tilde \gamma   E_1,  \\
    i\partial_t E_2 + E_1 +  \frac{ 2 \mathcal R  |E_2|^2}{|E_1|^2+|E_2|^2} E_2 &= -i  \tilde \gamma  E_2.
    \end{align*}
   Here  $E_1$ and $E_2$  are the amplitudes of the orthogonally polarised modes; 
   $\mathcal R>0$ is the nonlinearity parameter and 
    $\tilde \gamma>0$ is the gain-loss  coefficient.   The components $E_1$ and $E_2$ 
   can be 
   obtained from solutions  of \eqref{A3A4} with $|u|^2+|v|^2= 2 \mathcal R$ and $\gamma= \tilde \gamma / \mathcal R$
   by means of the transformation \eqref{E1E2}.

The cross-stimulated dimer  \eqref{A3A4} can be written in terms of  the spin variables \eqref{Stokes-variables}:
\begin{subequations}  
\label{32nd}
\begin{align}
    \dot{X} &= -YZ - \gamma X Z \label{A72}, \\
    \dot{Y} &= -Z + XZ - \gamma Y Z \label{A82}, \\
    \dot{Z} &= Y + \gamma (R^2 - Z^2) \label{A92}.
\end{align}
\end{subequations}
Here $X,Y,Z$ are functions of $\tau = 2t$, and overdots 
denote derivatives w.r.t $\tau$. 
The spin vector ${\bf S}= \{X,Y,Z\}$ will provide an appropriate basis for visualising the solutions of the Kerr dimer.

We further observe that 
the system \eqref{32nd} arises in the mean-field approximation of the many-body Bose-Hubbard model \cite{Graefe}. 

\subsection{Hamiltonian structure and integrability} 

If we define the polar coordinates $\eta$ and $\theta$
 such that
\begin{equation} 
X = \frac{1}{1+\gamma^2} + e^{-\eta}\cos\theta, 
\quad Y = -\frac{\gamma}{1+\gamma^2} + e^{-\eta}\sin\theta,
\label{pola}
\end{equation}
equations \eqref{A72} and \eqref{A82} reduce to
\be
\dot{\eta}  = \gamma Z, \quad \dot{\theta} = Z.  
\label{NewVars1}
\ee
Accordingly, 
\be
\xi = \eta - \gamma\theta- \gamma \arctan \gamma
\label{cq}
\ee
 is a conserved quantity,
in addition
to
\begin{equation}
    \mathcal R^2  = e^{-2\eta}  + Z^2
    + 2e^{-\eta}            \frac{\cos\theta - \gamma\sin\theta}{1+\gamma^2} + \frac{1}{1+\gamma^2}.
	\label{HH}
\end{equation}

To uncover the Hamiltonian formulation, 
we appoint $\eta$ and $\xi$ as two canonical coordinates
and the first integral 
\be
\mathcal H= \mathcal R^2 -\frac{1}{1+\gamma^2}
\label{HR}
\ee
 as the Hamiltonian:
\begin{align}
\mathcal{H}(\eta, \xi, p_\eta)   =   e^{-2\eta}  + \frac{\gamma^2}{4}p_\eta^2
  + \frac{2e^{-\eta}}{\sqrt{1+\gamma^2 }} 
    \cos    \left( \frac{ \eta - \xi}{\gamma} 
    \right).
    \label{HHH}
\end{align}
Here
$p_\eta= 2  \gamma^{-1}  Z$. 
The first equation in \eqref{NewVars1} and
equation \eqref{A92} acquire the form 
\[
\dot \eta= \frac{\partial \mathcal H}{\partial p_\eta},
\quad
\dot p_\eta= - \frac{\partial \mathcal H}{\partial \eta};
\]
hence $p_\eta$  is the momentum  canonically conjugate to $\eta$. 
The momentum $p_\xi$  conjugate to the coordinate $\xi$ can be found from the Hamilton equation
\[
\dot{p_\xi} = -\frac{\partial \mathcal{H}}{\partial \xi}, 
    \]
by integration:
\[
p_\xi = -\frac{2}{\gamma (1+\gamma^2)} 
\int_0^\tau 
\left[ \gamma X(\tau')+  Y(\tau') \right]  d\tau'.
    \]
    
   The existence of the canonical formulation 
   and an additional first integral ($\xi$) 
   establishes the Liouville integrability of the 
   dimer \eqref{A3A4}.
   
   \subsection{Fictitious particle formalism} 
   
 Transforming to the canonical variables allows us to obtain the general analytical solution
 of the system \eqref{32nd}.
 
Projections of trajectories 
 on the $(\theta, \dot \theta)$ plane are described by equations \eqref{HR}-\eqref{HHH}: 
\be
  {\dot \theta}^2+ U_\xi(\theta)
  = {\mathcal R}^2, 
    \label{PhasPor}
\ee
where
\begin{subequations}\label{J8} 
\be
U_\xi= \frac{1}{1+\gamma^2} 
  + e^{-2\eta}  
  +  \frac{2  e^{-\eta}
    \cos  ( \theta + \beta)}    {\sqrt{1+\gamma^2}},    \label{Ux}
    \ee
  \be
  \eta= \gamma (  \theta + \beta)  +\xi,         \label{J22} 
  \ee
  \end{subequations} 
  and 
  \[
  \beta= \arctan \gamma.
  \]
 Equation \eqref{PhasPor} can be interpreted as the energy conservation law for a fictitious Newtonian particle moving in a potential $U_\xi(\theta)$.
 The characterisation of the potential $U_\xi(\theta)$ will play  the key role  in the trajectory analysis. 
 
\begin{figure}[t]
 \begin{center} 
    \includegraphics*[width=\linewidth] {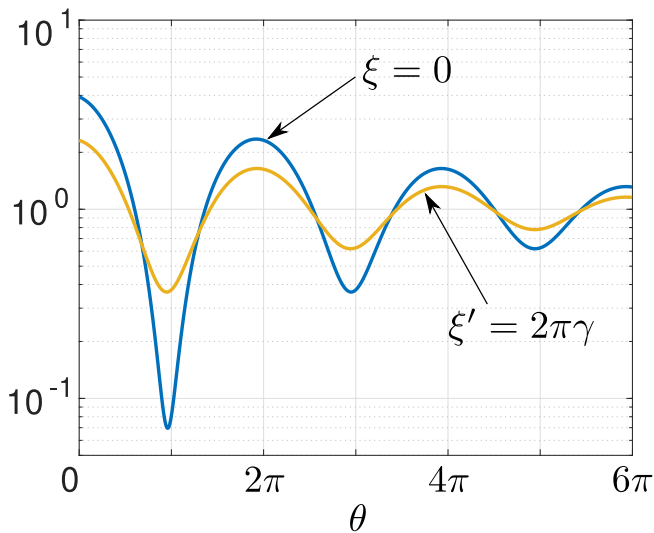} 
          \end{center}
  \caption{The potential $U_\xi(\theta)$ (blue) and its ``sibling" $U_{\xi^\prime}(\theta)$ (brown curve), with $\xi^\prime=\xi+ 2 \pi \gamma$.
 (In this plot,  $\gamma=\frac{1}{10}$,
  $\xi=0$,  and $\xi^\prime = \frac{2}{10}  \pi$.) The function $U_{\xi^\prime}(\theta)$  is obtained from  $U_\xi(\theta)$   by the  $2\pi$ horizontal translation of the latter. 
   \label{Ut}
  }
 \end{figure}

 We start by considering the interval $0 \leq \theta < 2 \pi$. 
 When the parameter $\xi$ satisfies
  $\xi <\xi_c$, where
\be
\label{xic}
\xi_c= -\frac{ 3 \pi}{2} \gamma + \ln \left( \gamma \sqrt{1+\gamma^2}\right),
\ee 
 the function $U_\xi(\theta)$  is monotonically decreasing in this interval. 
 As  $\xi$ is raised through $\xi_c$,  a pair of extrema  is born in  $[0, 2 \pi)$: a minimum at $\theta^{(1)}(\xi)$ and a maximum 
at $\theta^{(2)}(\xi)$, with $\theta^{(1)}<\theta^{(2)}$ (Fig   \ref{Ut}).  
The extrema are roots of the transcendental equation 
\be
\label{J12}
\sin(\theta+ 2 \beta)= - \gamma e^{-\eta}
\ee
with $\eta$ as in \eqref{J22};
hence $\theta^{(1,2)}$ satisfy
\be
\label{J14}
\sin(\theta^{(n)} +  2\beta) <0 \quad (n=1,2). 
\ee
The values of the potential at the extrema are
\be
U_\xi \left(\theta^{(n)} \right)= \frac{1}{\gamma^2} \sin^2 \left( \theta^{(n)} + \beta \right).
\label{J10}
\ee
At the bifurcation point, we have $\theta^{(1,2)}(\xi_c)= 3 \pi/2-\beta$ and 
so 
\be
\label{exta}
U_{\xi_c} (\theta^{(n)}(\xi_c))= \frac{1}{\gamma^2} \quad (n=1,2).
\ee

We also note an expression for the second derivative,
\be
\label{J15}
\left.  \frac{\partial^2 U_\xi}{\partial \theta^2} \right|_{\theta^{(n)}}= 
 \frac{ 2 \sin(\theta^{(n)} + 2\beta) \cos  (\theta^{(n)} +  \beta)}  {\sin \beta}.
\ee
Equation \eqref{J15} and  inequality \eqref{J14} imply that the point of minimum satisfies 
$
\cos ( \theta^{(1)} +\beta) <0
$
and the point of maximum has 
\be 
\cos (\theta^{(2)}+ \beta)>0.         
 \label{J17} 
\ee
With the help of \eqref{J14} and \eqref{J17},  simple trigonometry gives
\be
\sin (\theta^{(2)}+\beta)<0.  
 \label{J20} 
\ee

Taking advantage of the symmetry
\be \label{J61}
U_\xi(\theta+ 2 \pi)= U_{\xi+ 2 \pi \gamma} (\theta),
\ee
we can extend our analysis beyond the interval $[0, 2 \pi)$. Assuming $\xi>\xi_c$,
the potential $U_\xi(\theta)$ has a pair of extrema 
in each interval $[2 \pi \ell,  2\pi(\ell +1))$  with $\ell=1, 2, ...$.
The minimum is at 
\[
\theta^{(2 \ell+1)}(\xi)=\theta^{(2\ell- 1)} (\xi+2 \pi \gamma) +2 \pi
\]
 and the maximum at 
 \be
 \theta^{(2\ell+2)}(\xi)= \theta^{(2\ell)} (\xi+ 2 \pi \gamma)  +2 \pi.
 \label{J64} \ee
The value of the potential $U_\xi$ at its minimum $\theta^{(2\ell+1)}$ is equal to the value of the potential $U_{\xi +2 \pi \gamma}$
at its own local minimum in the interval $[2 \pi (\ell-1),  2 \pi \ell)$.  (See Fig \ref{Ut}.) 
A similar rule governs the local maxima: 
\be \label{J67}
U_\xi(\theta^{(2 \ell+2)}(\xi) ) = U_{\xi+2 \pi \gamma}(\theta^{(2\ell)}(\xi + 2\pi \gamma) ).
\ee

As the parameter $\xi$ is increased, the value of the potential $U_\xi$ at its maximum in $[0,  2 \pi)$   
decreases: 
\[
\frac{\mathrm d} {\mathrm d \xi} 
U_\xi   (\theta^{(2)}( \xi))=- \frac{2 \sin (\theta^{(2)}+\beta) \sin (\theta^{(2)}+2 \beta)}{\gamma^2 \sqrt{1+\gamma^2}}
<0.
\]
Here we took into account \eqref{J14}   and \eqref{J20}. 
We also note that according to \eqref{J12}, the point $\theta^{(1)}(\xi)$ approaches $\pi-2 \beta$ 
and $\theta^{(2)}(\xi)$  approaches $2\pi-2\beta$
as $\xi \to \infty$. 
Hence, by equation \eqref{J10}
the maximum value of the potential in the interval $[0, 2\pi)$ is bounded from below:
\be
U_\xi(\theta^{(2)}(\xi)) > \frac{1}{1+\gamma^2}.
\label{J30}
\ee

Finally, the inequality \eqref{J14} implies
\be
\cos (\theta^{(2)}+\beta) < \cos \beta.
\label{J52}
\ee
Making use of \eqref{J52} one can establish the following  relation, valid for all $\xi$:
\[
U_\xi(0) > U_\xi(\theta^{(2)}(\xi)).
\]
The symmetry identities \eqref{J61}, \eqref{J64} and \eqref{J67} yield the inequality
\be 
\label{J54}
U_\xi(2\pi) > U_\xi(\theta^{(4)}(\xi)).
\ee

\subsection{Spin trajectories      from    particle flight paths} 

Consider a  trajectory   of the spin system \eqref{32nd}   passing through a point on the equator ($Z=0$) at time $\tau=0$. 
The corresponding fictitious particle starts its motion from rest ($ \dot \theta(0)=0$), with the values of integrals  $\xi$ and
$\mathcal R$ defined by
the initial data, $\theta_0$ and $\eta(0)$:
\[
\xi= \eta(0)-\gamma (\theta_0 +\beta), \quad
{\mathcal R}=  \sqrt{
U_\xi(\theta_0)}.
\]
The 
implicit solution $\tau(\theta)$ of equation \eqref{PhasPor} is given
by the integral
\be
 \tau = \pm 
\int_{\theta_0}^\theta 
  \frac{   d \theta'}{\sqrt{
 {\mathcal R}^2- U_\xi(\theta')
      }}.
    \label{J7}
\ee
The solution is valid for all $\theta$ such that the expression under the radical is positive in the interval $(\theta_0, \theta)$.
Without loss of generality we may let $\theta_0$ lie in the interval $[0, 2\pi)$.

If $\partial U_\xi/ \partial \theta<0$ at $\theta=\theta_0$, the $\theta$-particle will start moving in the positive direction 
and we choose the positive sign in \eqref{J7}. 
The corresponding spin trajectory ${\bf S}(\tau)$ will emerge into the northern hemisphere. 
If $\partial U_\xi/ \partial \theta>0$ at $\theta=\theta_0$, the particle will start moving in the  negative direction 
and we choose the negative sign in \eqref{J7}. In that case, the point ${\bf S}$ will move into the southern hemisphere.

The negative-time motions can be classified in a similar manner.

For a given value of $\gamma$,  the character of motion
is determined 
by the values of the parameters $\mathcal R$ and $\xi$. 
Since two or three values of $\theta_0 \in [0, 2\pi)$ can be mapped to the same $U_\xi$, we also need to indicate the position of $\theta_0$ 
relative to the minimum and maximum of $U_\xi(\theta)$.

 {\bf (a)} Assume first that $\mathcal R > \gamma^{-1}$. 
 If $\xi> \xi_c$,  the potential $U_\xi(\theta)$
 has a sequence of local maxima  at $\theta^{(2 \ell)}$, $\ell=1,2,...$,
 while if 
 $ \xi_c-2\pi \gamma  m < \xi < \xi_c- 2 \pi \gamma (m-1)$
($m=1,2, ...$), 
  the potential  is monotonically decreasing in $(-\infty, 2 \pi m)$
  but 
  has local maxima in each interval
$[2 \pi \ell,  2\pi(\ell +1))$ with $\ell=m,m+1, ...$. 
 In either case, 
 all local maxima $U_\xi(\theta^{(2\ell)})$ lie below $\gamma^{-2}$ (cf. \eqref{J10}). 
Consequently, the particle will accelerate to some positive speed and then continue moving  with an oscillatory positive
  velocity  bounded from below:   $\dot \theta> \sqrt{{\mathcal R}^2-\gamma^{-2}}$. Regardless of $\xi$,  the particle will eventually escape to infinity:  $\theta \to \infty$,  with $\eta \to \infty$ as well. 
 A similar asymptotic behaviour occurs in the negative-time domain: $\eta \to \infty$ as $\tau \to -\infty$.

\begin{figure}[t]
 \begin{center} 
    \includegraphics*[width=0.65\linewidth] {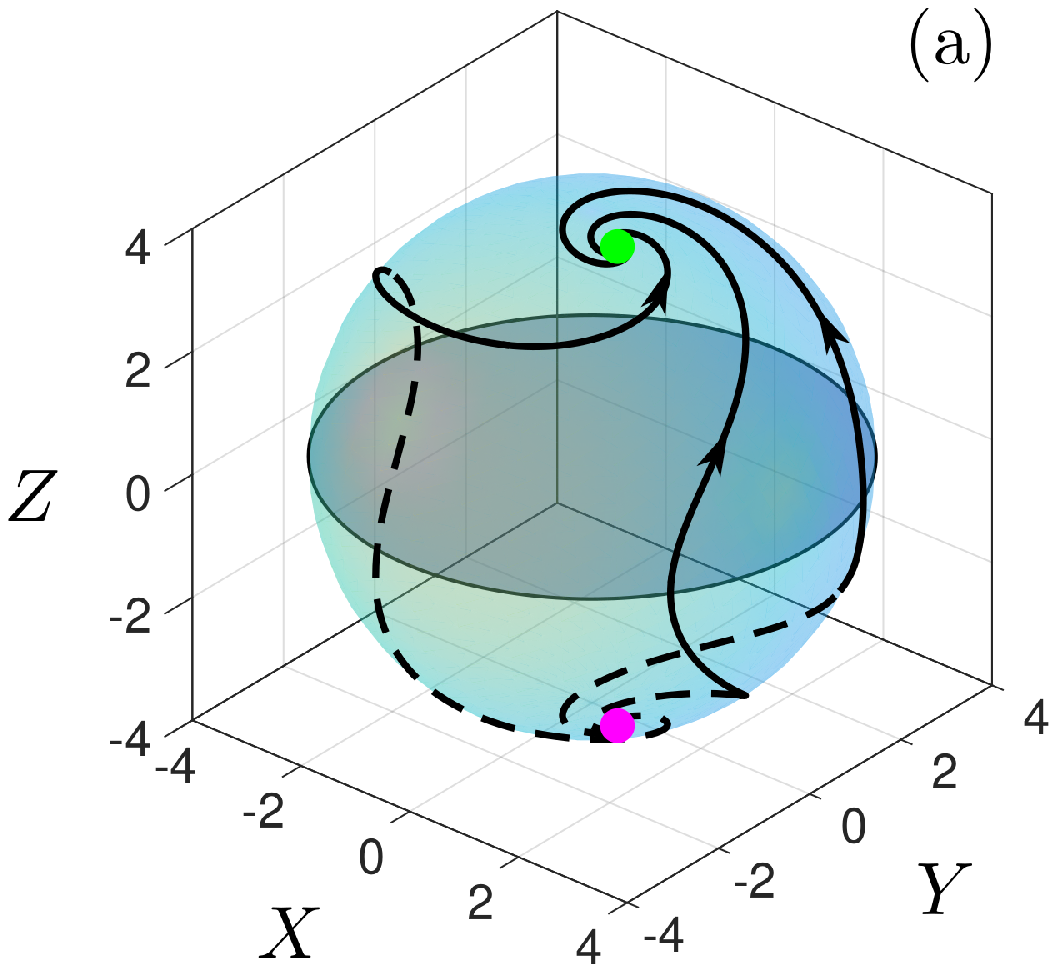}  
       \includegraphics*[width=0.65\linewidth] {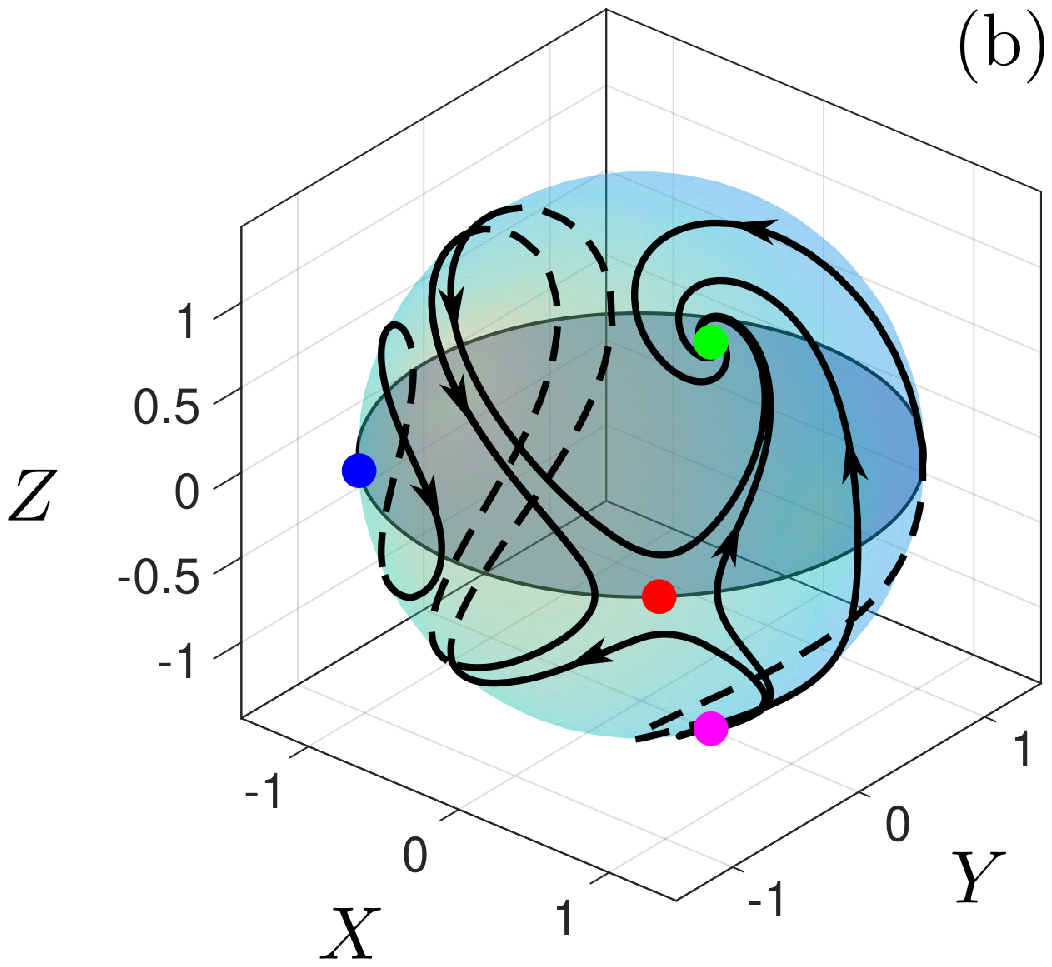}   
          \end{center}
  \caption{Trajectories of the system \eqref{32nd}  on the surface of the sphere of the 
  radius  $\mathcal R>  \gamma^{-1}$   (a) and    $ (1+\gamma^2)^{-1/2}  <\mathcal R< \gamma^{-1}$
  (b). In this and the next figure,  $\gamma=\frac12$. 
   \label{largeR}
  }
 \end{figure}

The implicit solution \eqref{J7}  with $\eta \to \infty$ as $\tau \to  \pm \infty$
admits a simple interpretation in terms of the  spin components \eqref{pola}.
 (The corresponding trajectories on the spin sphere have been  numerically delineated  in  Ref \cite{Graefe}.)
 Similar to 
the system \eqref{1st3} (Fig \ref{3figs} (a))
the sphere with $\mathcal R > \gamma^{-1}$ supports a pair of latitudinal fixed points,
with 
\be
\label{J1} 
X=\frac{1}{1+\gamma^2}, \quad 
Y=-  \gamma X, \quad  
Z= \pm \sqrt{{\mathcal R}^2-X}.
\ee
The fictitious particle's journeys from infinity   to $\theta_0$  and back to infinity correspond to
trajectories ${\bf S}(\tau)$ emerging from  the unstable focus in the southern hemisphere and spiralling into the attractor in the north (Fig \ref{largeR}(a)).

{\bf (b)} The next range to consider is  $(1+\gamma^2)^{-1/2}  < \mathcal R   < \gamma^{-1}$.
Let  the function $\xi_d(\mathcal R)$ be defined as 
the inverse  of the 
monotonically decreasing function ${\mathcal R}=\sqrt{ U_\xi(\theta^{(2)}(\xi))}$.
In view of \eqref{exta}, 
  $\xi_d (\mathcal R) >\xi_c$ for all $\mathcal R$ in the current range. 
  As $\mathcal R$ approaches  $\gamma^{-1}$ from below,    the value $\xi_d (\mathcal R)$ approaches  $ \xi_c$; as $\mathcal R$ approaches  $(1+\gamma^2)^{-1/2} $ from above,   we have 
  $\xi_d \to \infty$.

When $\xi< \xi_c$  or
 $\xi > 
\xi_d(\mathcal R)$, the total energy ${\mathcal R}^2$ of the particle is greater than $U_\xi$ at all of its local maxima. 
Accordingly, 
 the particle escapes to infinity as $\tau \to \pm \infty$.     This class of motions corresponds to the heteroclinic trajectories  on the sphere
 flowing from the southern to the northern focus in \eqref{J1} (Fig \ref{largeR} (b)).

  Turning to the interval
 $\xi_c< \xi< \xi_d$ we first assume that the point $\theta_0$ lies to the left of the local maximum $\theta^{(2)}$. 
 In that case
  the energy of the $\theta$-particle is insufficient to overcome the potential barrier: ${\mathcal R}^2 < U_\xi(\theta^{(2)} (\xi))$.
 The  particle becomes trapped in the potential well, and  its motion is periodic with  period
 \[
 T=   \left|  \int_{\theta_0}^{\theta_{\mathrm{turn}}} \frac{d \theta'}
 {\sqrt{{\mathcal R}^2-U_\xi(\theta')}} \right|.
 \]
 Here $\theta_{\mathrm{turn}}$ is the second root of the equation ${\mathcal R}^2=U_\xi(\theta)$ to the left of the maximum of the 
 potential in the interval $[0, 2\pi)$.

To interpret the periodic motions in terms of trajectories on the sphere, we note that 
as $\mathcal R$ is decreased through $\gamma^{-1}$, 
a saddle-centre bifurcation  brings about
two new fixed points lying on the equator:
\be
\label{J2}
X= \pm \mathcal R \sqrt{1-\gamma^2 {\mathcal R}^2}, 
\quad
Y=-\gamma {\mathcal R}^2,
\quad Z=0.
\ee
For $\mathcal R$  in the interval 
$(1+\gamma^2)^{-1/2}   < \mathcal R < \gamma^{-1}$, 
the point with negative $X$ is a centre and the one with positive $X$ is a saddle.
The oscillations of the $\theta$-particle in the potential well 
translate into a thicket of closed orbits surrounding the centre
  (Fig \ref{largeR}(b)).
  
  If the point $\theta_0$ lies to the right of the local maximum $\theta^{(2)}$ (that is, if 
  $\theta^{(2)}< \theta_0< 2 \pi$), the particle will escape to infinity: $\theta \to \infty$ as $\tau \to \infty$. 
  It cannot be captured by the potential well in the interval $[2\pi, 4\pi)$ because $U_\xi(\theta_0)$
  is greater than  $U_\xi(2\pi)$ and therefore, by  inequality \eqref{J54}, greater than the potential barrier
  $U_\xi(\theta^{(4)}(\xi))$.    

By examining the neighbourhood of the saddle point in Fig \ref{largeR}(b)), one can readily reconstruct
a homoclinic curve connecting the saddle to itself. This trajectory
results by choosing
$\xi=\xi_d(\mathcal R)$.
The homoclinic curve separates 
the family of closed orbits 
  from the focus-to-focus flows.

 {\bf (c)} Finally, it remains to examine the range $\mathcal R < (1+\gamma^2)^{-1/2}$. 
 Since the function $U_\xi(\theta)$ with $\xi< \xi_c$ is monotonically decreasing in $[0, 2\pi)$,
 and since $U_\xi(2\pi) > (1+\gamma^2)^{-1}$ by equation \eqref{Ux},
  this  parameter range  is only accessible to initial conditions with $\xi  > \xi_c$.
In view of \eqref{J30}, the $\theta$-particle  with $\dot \theta(0)=0$ and $\theta_0$ 
 satisfying $U_\xi(\theta_0)={\mathcal R}^2$ 
finds itself trapped in a potential well. 

\begin{figure}[t]
 \begin{center} 
        \includegraphics*[width=0.65\linewidth] {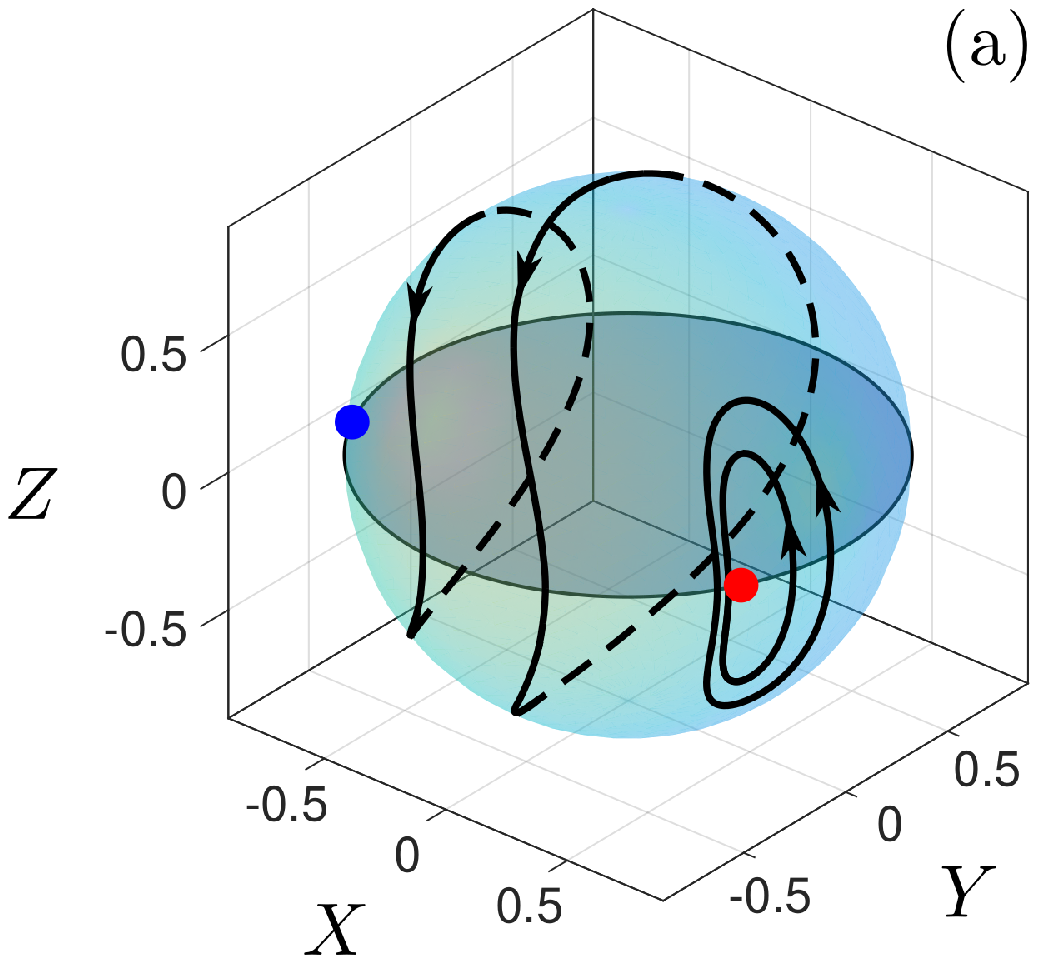}  
                     \includegraphics*[width=0.65\linewidth] {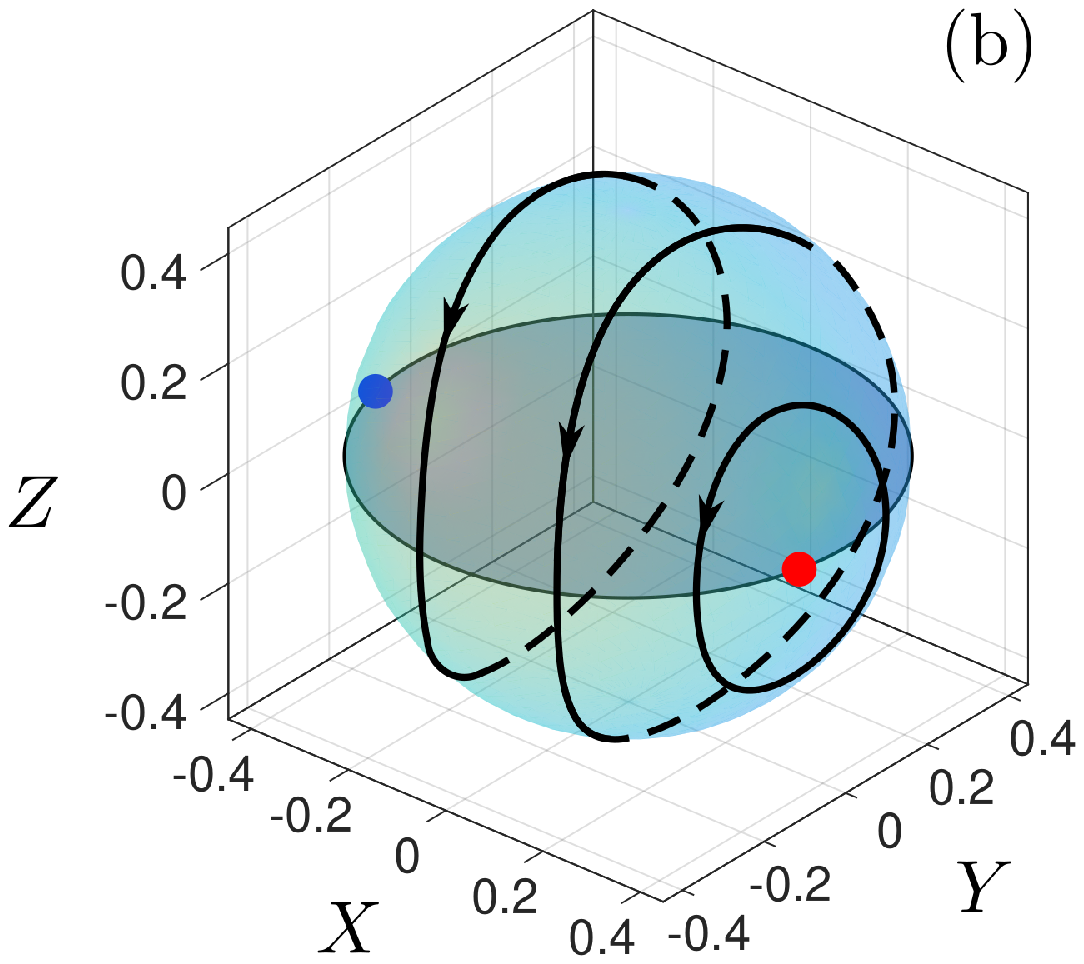}          
          \end{center}
  \caption{Trajectories of the system \eqref{32nd} on the surface of the sphere of the 
  radius
 $\mathcal R= \frac{1}{\sqrt{1+\gamma^2}}$  (a) and
     $\mathcal R < \frac{1}{\sqrt{1+\gamma^2}}$ 
 (b).
 \label{smallR}
  }
 \end{figure}

The disappearance  of aperiodic solutions $\theta(\tau)$  
is consistent with the bifurcation occurring as $\mathcal R$ is reduced  below $(1+\gamma^2)^{-1/2}$.
At this value of $\mathcal R$, 
  the pair of latitudinal fixed points \eqref{J1} merges with the saddle on the equator forming
  the second centre (Fig \ref{smallR}(a)). The two centre points are given by equations 
\eqref{J2}; each point is surrounded by a family of closed orbits (Fig \ref{smallR} (b)). 
The closed  orbits  are described by   periodic motions of the  particle in the potential well.


\section{Integrable \PT-symmetric  necklaces}
\label{necks}

 In a nonhermitian necklace of more than two waveguides, the trapping of trajectories  in a finite part of the phase space 
 may require each site to coordinate its gain and loss  rate with both of its left and right neighbours. 
 This  {\it cross-compensation\/} mechanism is more subtle than the cross-stimulation of two channels of a dimer. 
 In this section,  we exemplify the cross-compensation with arrays consisting of three and four elements. 

\subsection{Trimer}

We identified several \PT-symmetric trimers endowed with a
Hamiltonian structure and possessing an integral of motion that prevents the blow-up behaviour. 
However, only one of these models has three first integrals in involution and defines a Liouville-integrable dynamical system.

The trimer in question is a nonhermitian extension of the closed Ablowitz-Ladik chain:
\begin{subequations}
  \label{AL} 
\begin{align}
    i u_t &= (1+|u|^2)(v+w)(1+i\gamma),                              \label{AL1} \\
    i v_t &= (1+|v|^2)\left[ u+w +i\gamma(w-u)\right],           \label{AL2}   \\
    i  w_t&= (1+|w|^2)(u+v)(1-i\gamma).                                    \label{AL3}
\end{align}
\end{subequations}
The system is invariant under the product of the $\mathcal P$ and $\mathcal T$ transformations, 
where
\[
\mathcal Pu(t)= w(t), \quad \mathcal P v(t) =v(t), \quad \mathcal P w(t)=u(t)
\]
and 
\[
\mathcal T u(t)= u^*(-t), 
\quad 
\mathcal T v(t)=v^*(-t),
\quad
\mathcal T w(t)=w^*(-t).
\]

Similar to the dimers \eqref{A100} and \eqref{A3A4}, 
the linearisation  of equations \eqref{AL} about  $u=v=w=0$ gives a system with a hermitian matrix and a purely real  spectrum.
Accordingly, the trimer  \eqref{AL} 
does not suffer the linearised  \PT-symmetry breaking as the value of $\gamma$ is raised.

The system \eqref{AL} admits a canonical representation 
\begin{equation}
     u_t= \{H,u\},
    \quad  v_t = \{H,v\}, 
    \quad  w_t = \{H,w\}\notag
\end{equation}
with the Hamilton function
\begin{equation}
    H = (1+i\gamma)({u^*}v+ {v^*}w + w{u^*} ) + \text{c.c.}      \label{Ham}
\end{equation}
and 
the Gerdjikov-Ivanov-Kulish (GIK) bracket \cite{GIK}. Here
\begin{gather*}
    \{u,   {u^*}\} = i(1+|u|^2),    \quad
    \{v,   {v^*}\} = i(1+|v|^2),   \\     
    \{w,  {w^*}\} = i(1+|w|^2),
    \end{gather*}
while all other brackets are equal to zero: $\{u,v \} = \{u, v^* \} = ...=0$.

In addition to the Hamiltonian, the system conserves the total
momentum,
\[
M = (\gamma -i)({u^*}v + {v^*}w -  w{u^*}) + \text{c.c.},
\]
and a product
\[
     \Pi  = (1+|u|^2)(1+|v|^2)(1+|w|^2).
    \]
The integrals  $M$   and $\Pi$
commute: $\{  M, \Pi \}=0$.
Consequently, the system \eqref{AL}  is completely integrable.

Note that  since the total power $P=|u|^2 + |v|^2+|w|^2$ is bounded  from above
by the  conserved quantity $\Pi$,
 the trimer does not exhibit any unbounded trajectories.

 It is worth noting that the absence of blow-up regimes is not 
related to the integrability of the model but is instead a consequence of the compensation of gain and loss rate in 
the neighbouring channels. This can be illustrated by the following family of nonintegrable cross-compensated trimers:
\begin{subequations}
  \label{ALf} 
\begin{align}
    i u_t &= f_1(|u|^2) (v+w)  (1+i\gamma),                          \\
    i v_t &=f_2(|v|^2)   \left[ u+w +i\gamma(w-u)\right], 
            \\
    i  w_t&= f_3(|w|^2)(u+v)(1-i\gamma).                       
\end{align}
\end{subequations} 
In  equations \eqref{ALf},  $f_n(\rho)$ ($n=1,2,3$) are positive functions with $f_n(0)=1$ and $f_n(\rho) \leq 1$ for $\rho>0$.
 The model \eqref{ALf} is Hamiltonian with the Hamilton function \eqref{Ham} and obvious modification
 of the bracket. It has the first integral
 \[
 I= F_1(|u|^2)+F_2(|v|^2) +F_3(|w|^2), 
 \]
 where 
 \[
 F_n(\rho)= \int_0^\rho \frac{d \rho'}{f_n(\rho')}.
 \]
  Since $F_n(\rho) \geq \rho$,
 the total power is bounded from above by $I$ and no trajectory can escape to infinity.

\subsection{Quadrimer}

Guided by  the symplectic structure and cross-compensating arrangement 
 of the integrable \PT-symmetric trimer,
it is not difficult to construct a four-waveguide necklace with similar properties:
\begin{subequations}
  \label{4AL} 
\begin{align}
    i \dot u_1 &= (1+|u_1|^2)(u_2+u_4)(1+i\gamma), \label{4AL1} \\
    i \dot u_2 &= (1+|u_2|^2)\left[ u_3+u_1 +i\gamma(u_3-u_1)\right], \label{4AL2} \\
    i  \dot u_3 &= (1+|u_3|^2)\left[ u_4+u_2 +i\gamma(u_4-u_2)\right], \label{4AL3}\\
    i \dot u_4 &= (1+|u_4|^2)(u_1+u_3)(1-i\gamma). \label{4AL4}
\end{align}
\end{subequations}
The quadrimer \eqref{4AL} is \PT-symmetric, with the $\mathcal P$ operator
defined by 
\begin{align*}
     \mathcal Pu_1(t)= u_4(t), \quad \mathcal P u_2(t) =u_3(t),  \\
     \mathcal Pu_3(t)= u_2(t), \quad \mathcal P u_4(t) =u_1(t),                  
\end{align*}
and $\mathcal T$ as in $\mathcal T u_n(t)= u_n^*(-t)$. 
The linearisation of the quadrimer \eqref{4AL} about $u_1=u_2=u_3=u_4=0$  gives a system with a hermitian matrix and a purely real spectrum.

The quadrimer retains a similar canonical representation to the trimer \eqref{AL}, with the Hamiltonian
\begin{equation}
    H = (1+i\gamma)({u_1^*}u_2+ {u_2^*}u_3 + {u_3^*}u_4 + u_4{u_1^*} ) + \text{c.c.}
\end{equation}
and GIK bracket. Here 
\begin{gather*}
    \{u_n,   {u_m^*}\} = i(1+|u_n|^2) \delta_{nm}, \\
       \{u_n,   {u_m}\}=0, \quad    \{u_n^*,   {u_m}^*\}=0.        
       \end{gather*}

 In direct analogy to the trimer, the system conserves the total momentum
\[
M = (\gamma -i)({u_1^*}u_2+ {u_2^*}u_3 + {u_3^*}u_4 - u_4 u_1^* ) + \text{c.c.},
\]
as well as the product
\[
     \Pi  = \prod_{n=1}^{4}(1+|u_n|^2).
    \]
    In addition, the 
    quadrimer  conserves a quartic quantity
\begin{equation}
    I = (1-i\gamma)(u_1  u_2^*  u_3 u_4^* -u_2 u_4^* - u_1 u_3^*) + c.c.
\end{equation}
All four integrals $H$, $M$, $\Pi$ and $I$ mutually commute, and consequently the system \eqref{4AL} 
is completely integrable. \\

\section{Concluding remarks}
\label{Conclusions}

In this paper, we explored  a class of the  \PT-symmetric discrete Schr\"odinger equations  with purely-nonlinear nonhermitian terms.
An a priori advantage of systems of this type in physics is that 
the trivial solution remains stable and  linearised \PT-symmetry  unbroken
regardless of the value of the  gain-loss coefficient.

We have identified  two  nonequivalent  dimers  that, in addition to preserving the linearised \PT-symmetry,    conserve  the quantity $|u|^2+|v|^2$.
(In the optical context, this means that the total power of light is conserved.)
The power conservation is brought about by the cross-stimulation of the two channels of the dimer.
The constancy of $|u|^2+|v|^2$ ensures that no trajectories of these dynamical systems 
escape to infinity.

Remarkably, both dimers possess a canonical structure and a first integral independent of the Hamiltonian.
This establishes the Liouville integrability of the two systems. The 
 transformation  to the canonical variables allowed us to construct their general analytical solution.

The cross-compensation of the gain and loss rate in the  neighbouring channels remains an efficient blow-up prevention 
mechanism  in the case of the nonhermitian Schr\"odinger {\it necklaces} --- the ring-shaped arrays  of $N$ waveguides \cite{necklaces}.
 We have   exemplified this idea with the construction of a
 family of   trimers  ($N=3$) whose trajectories are confined to a finite part of their phase space.
 The entire family is endowed with a canonical structure while one member of the family has
  3 first integrals in involution and defines a completely integrable system. 
 
 Finally,  we have identified a completely integrable cross-compensated quadrimer --- a  \PT-symmetric necklace of $N=4$ waveguides.

\acknowledgements
A discussion with  Robert McKay is 
 gratefully acknowledged. 
We thank Andrey Miroshnichenko and Tsampikos Kottos for instructive correspondence. 
This research was supported by the NRF of South Africa (grant No 120844).

\end{document}